\documentclass[12pt]{article}
\usepackage{epsfig}
\usepackage{amsmath}
\usepackage{hhline}
\usepackage{amssymb}
\usepackage{times}

\newlength{\dinwidth}
\newlength{\dinmargin}
\begin{document}  
% The rest
\newcommand{\pom}{{I\!\!P}}
\newcommand{\reg}{{I\!\!R}}
\newcommand{\slowpi}{\pi_{\mathit{slow}}}
\newcommand{\fiidiii}{F_2^{D(3)}}
\newcommand{\fiidiiiarg}{\fiidiii\,(\beta,\,Q^2,\,x)}
\newcommand{\n}{1.19\pm 0.06 (stat.) \pm0.07 (syst.)}
\newcommand{\nz}{1.30\pm 0.08 (stat.)^{+0.08}_{-0.14} (syst.)}
\newcommand{\fiidiiiful}{F_2^{D(4)}\,(\beta,\,Q^2,\,x,\,t)}
\newcommand{\fiipom}{\tilde F_2^D}
\newcommand{\ALPHA}{1.10\pm0.03 (stat.) \pm0.04 (syst.)}
\newcommand{\ALPHAZ}{1.15\pm0.04 (stat.)^{+0.04}_{-0.07} (syst.)}
\newcommand{\fiipomarg}{\fiipom\,(\beta,\,Q^2)}
\newcommand{\pomflux}{f_{\pom / p}}
\newcommand{\nxpom}{1.19\pm 0.06 (stat.) \pm0.07 (syst.)}
\newcommand {\gapprox}
   {\raisebox{-0.7ex}{$\stackrel {\textstyle>}{\sim}$}}
\newcommand {\lapprox}
   {\raisebox{-0.7ex}{$\stackrel {\textstyle<}{\sim}$}}
\def\gsim{\,\lower.25ex\hbox{$\scriptstyle\sim$}\kern-1.30ex%
\raise 0.55ex\hbox{$\scriptstyle >$}\,}
\def\lsim{\,\lower.25ex\hbox{$\scriptstyle\sim$}\kern-1.30ex%
\raise 0.55ex\hbox{$\scriptstyle <$}\,}
\newcommand{\pomfluxarg}{f_{\pom / p}\,(x_\pom)}
\newcommand{\dsf}{\mbox{$F_2^{D(3)}$}}
\newcommand{\dsfva}{\mbox{$F_2^{D(3)}(\beta,Q^2,x_{I\!\!P})$}}
\newcommand{\dsfvb}{\mbox{$F_2^{D(3)}(\beta,Q^2,x)$}}
\newcommand{\dsfpom}{$F_2^{I\!\!P}$}
\newcommand{\gap}{\stackrel{>}{\sim}}
\newcommand{\lap}{\stackrel{<}{\sim}}
\newcommand{\fem}{$F_2^{em}$}
\newcommand{\tsnmp}{$\tilde{\sigma}_{NC}(e^{\mp})$}
\newcommand{\tsnm}{$\tilde{\sigma}_{NC}(e^-)$}
\newcommand{\tsnp}{$\tilde{\sigma}_{NC}(e^+)$}
\newcommand{\st}{$\star$}
\newcommand{\sst}{$\star \star$}
\newcommand{\ssst}{$\star \star \star$}
\newcommand{\sssst}{$\star \star \star \star$}
\newcommand{\tw}{\theta_W}
\newcommand{\sw}{\sin{\theta_W}}
\newcommand{\cw}{\cos{\theta_W}}
\newcommand{\sww}{\sin^2{\theta_W}}
\newcommand{\cww}{\cos^2{\theta_W}}
\newcommand{\trm}{m_{\perp}}
\newcommand{\trp}{p_{\perp}}
\newcommand{\trmm}{m_{\perp}^2}
\newcommand{\trpp}{p_{\perp}^2}
\newcommand{\alp}{\alpha_s}

\newcommand{\alps}{\alpha_s}
\newcommand{\sqrts}{$\sqrt{s}$}
\newcommand{\LO}{$O(\alpha_s^0)$}
\newcommand{\Oa}{$O(\alpha_s)$}
\newcommand{\Oaa}{$O(\alpha_s^2)$}
\newcommand{\PT}{p_{\perp}}
\newcommand{\JPSI}{J/\psi}
\newcommand{\sh}{\hat{s}}
\newcommand{\uh}{\hat{u}}
\newcommand{\MP}{m_{J/\psi}}
\newcommand{\PO}{I\!\!P}
\newcommand{\xpom}{x_{\PO}}
\newcommand{\ttbs}{\char'134}
\newcommand{\xpomlo}{3\times10^{-4}}  
\newcommand{\xpomup}{0.05}  
\newcommand{\dgr}{^\circ}
\newcommand{\pbarnt}{\,\mbox{{\rm pb$^{-1}$}}}
\newcommand{\gev}{\,\mbox{GeV}}
\newcommand{\WBoson}{\mbox{$W$}}
\newcommand{\fbarn}{\,\mbox{{\rm fb}}}
\newcommand{\fbarnt}{\,\mbox{{\rm fb$^{-1}$}}}
%
% Some useful tex commands
%
\newcommand{\xbj}{\ensuremath{x}}
\newcommand{\ybj}{\ensuremath{y}}
\newcommand{\qsq}{\ensuremath{Q^2} }
\newcommand{\gevsq}{\ensuremath{\mathrm{GeV}^2} }
\newcommand{\et}{\ensuremath{E_t^*} }
\newcommand{\rap}{\ensuremath{\eta^*} }
\newcommand{\gp}{\ensuremath{\gamma^*}p }
\newcommand{\dsiget}{\ensuremath{{\rm d}\sigma_{ep}/{\rm d}E_t^*} }
\newcommand{\dsigrap}{\ensuremath{{\rm d}\sigma_{ep}/{\rm d}\eta^*} }
%
% Bibliography-style
%
\bibliographystyle{pi096}
%
% definitions specific for this paper
%
\newcommand{\lum}{\ensuremath{\cal L}}
\newcommand{\piz}{\ensuremath{\pi^{\circ}}}
\newcommand{\ptpi}{\ensuremath{p_{T,\pi}^{\star}}}
\newcommand{\etapi}{\ensuremath{\eta_{\pi}}}
\newcommand{\thpi}{\ensuremath{\theta_{\pi}}}
\newcommand{\pt}{\ensuremath{p_{T}}}
\newcommand{\xpi}{\ensuremath{x_{\pi}}}
\newcommand{\epi}{\ensuremath{E_{\pi}}}
\newcommand{\eprot}{\ensuremath{E_{proton}}}
%
% Journal macro
%
\def\Journal#1#2#3#4{{#1} {\bf #2} (#3) #4}
\def\NCA{\em Nuovo Cimento}
\def\NIM{\em Nucl. Instrum. Methods}
\def\NIMA{{\em Nucl. Instrum. Methods} {\bf A}}
\def\NPB{{\em Nucl. Phys.}   {\bf B}}
\def\PLB{{\em Phys. Lett.}   {\bf B}}
\def\PRL{\em Phys. Rev. Lett.}
\def\PRD{{\em Phys. Rev.}    {\bf D}}
\def\ZPC{{\em Z. Phys.}      {\bf C}}
\def\EJC{{\em Eur. Phys. J.} {\bf C}}
\def\CPC{\em Comp. Phys. Commun.}

\begin{titlepage}

\noindent
DESY 99-094 \hfill ISSN 0418-9833\\
July 1999

\vspace{2cm}

\begin{center}
\begin{Large}

{\bf Forward \boldmath {\piz}-Meson Production \unboldmath at HERA}

\vspace{2cm}

H1 Collaboration

\end{Large}
\end{center}

\vspace{2cm}

\begin{abstract}
\noindent
 High transverse momentum {\piz}-mesons have been measured
with the H1 detector at HERA
 in deep-inelastic $ep$ scattering events at
 low Bjorken-$x$, down to  $x \approx 4{\cdot}10^{-5}$.
 The measurement is performed 
 in a region of small angles with respect to  the
 proton remnant in the laboratory frame of reference, namely the  
 forward region, and 
 corresponds to central
 rapidity in the centre of mass system of the virtual photon and proton.
 This region is expected to be   
 particularly sensitive to QCD  effects in hadronic final
 states. Differential cross-sections 
 for inclusive {\piz}-meson production 
are presented as a function of Bjorken-$x$ and
 the four-momentum transfer $Q^2$, and  as a function of
 transverse momentum and pseudorapidity. % of the {\piz}-mesons.
 A recent numerical BFKL calculation and predictions from QCD models based 
 on  DGLAP
 parton evolution are compared with  the data.

\end{abstract}

\vspace{1.5cm}

\begin{center}
To be submitted to Phys. Lett. 
\end{center}

\end{titlepage}

\begin{Large} \begin{center} H1 Collaboration \end{center} \end{Large}
\begin{flushleft}
%   H1AUTS  Author list by names, no. of authors  344
%           status: 04/07/99   15.20.52
 C.~Adloff$^{33}$,                %WUPP-ST                  Adloff             
 V.~Andreev$^{24}$,               %LPI -PD                  Andreev            
 B.~Andrieu$^{27}$,               %ECPL-PD                  Andrieu            
 V.~Arkadov$^{34}$,               %ZEUT-PD    10/96         Arkadov            
 A.~Astvatsatourov$^{34}$,        %ZEUT-ST     2/98         Astvatsatourov     
 I.~Ayyaz$^{28}$,                 %PARI-ST       5/96       Ayyaz              
 A.~Babaev$^{23}$,                %ITEP-PD                  Babaev             
 J.~B\"ahr$^{34}$,                %ZEUT-PD                  Baehr              
 P.~Baranov$^{24}$,               %LPI -PD                  Baranovp           
 E.~Barrelet$^{28}$
% ...

\end{flushleft}
\begin{flushleft} {\it
%     H1 Institutes as appearing on publications
% ...
$ ^{23}$ Institute for Theoretical and Experimental Physics,
          Moscow, Russia \\
 $ ^{24}$ Lebedev Physical Institute, Moscow, Russia$^{f,k}$ \\
% ...
$ ^{27}$ LPNHE, \'{E}cole Polytechnique, IN2P3-CNRS, Palaiseau, France \\
 $ ^{28}$ LPNHE, Universit\'{e}s Paris VI and VII, IN2P3-CNRS,
          Paris, France \\
% ...
$ ^{33}$ Fachbereich Physik, Bergische Universit\"at Gesamthochschule
          Wuppertal, Wuppertal, Germany$^a$ \\
 $ ^{34}$ DESY, Zeuthen, Germany$^a$ \\
% ...
$ ^a$ Supported by the Bundesministerium f\"ur Bildung, Wissenschaft,
        Forschung und Technologie, FRG,
        under contract numbers 7AC17P, 7AC47P, 7DO55P, 7HH17I, 7HH27P,
        7HD17P, 7HD27P, 7KI17I, 6MP17I and 7WT87P \\
% ...
$ ^f$ Supported by the Deutsche Forschungsgemeinschaft \\
% ...
$ ^k$ Supported by Russian Foundation for Basic Research 
       grant no. 96-02-00019 \\
% ...
% $ ^{m}$ Foundation for Polish Science fellow \\

   } \end{flushleft}

\newpage

\section{Introduction}
\noindent
Hadronic final state analyses in Deep-Inelastic Scattering (DIS) interactions
at HERA allow novel 
stringent
tests of the physics of Quantum Chromodynamics (QCD), the theory 
of the strong interactions, in a kinematical region of high parton
densities which so far has not been 
accessible~\cite{mueller,dhotref,hotref,KwMaOu}. The high Center of Mass System
(CMS)  energy 
($\sim$ 300 GeV) of the HERA collider allows  a region in 
Bjorken-$x$ of $\sim 10^{-5}-10^{-4}$ to be reached 
while keeping the momentum transfer, $Q^2$,
larger than a few GeV$^2$, hence remaining in the regime of perturbative
QCD (pQCD). In  DIS a parton in the proton can induce a QCD
cascade consisting of several subsequent parton emissions before the
final parton interacts with the virtual photon.
The multiplicity and the $x$
distribution of these emitted partons differ significantly in different
approximations of QCD dynamics at small {\xbj}.

At low $x$, pQCD evolution is complicated by the occurrence of two 
large logarithms in the evolution equations, 
namely $\ln 1/x$ and $\ln Q^2$.  In contrast, in the better tested 
 region of pQCD at larger $x$  a summation of the
leading $\ln Q^2$ terms is sufficient. A complete
perturbative treatment in the low-$x$ region is not yet available, and
different approximations are made resulting in
different parton dynamics. 
 At high $Q^2$ and high $x$ pQCD requires  
 the resummation of contributions of $\alpha_s\ln(Q^2/Q_0^2)$ terms, yielding 
the DGLAP (Dokshitzer-Gribov-Lipatov-Altarelli-Parisi)~\cite{dglap}
evolution equations. However at small $x$ the contribution of large
 leading $\ln 1/x$ terms may become important. Resummation of these
terms leads to the BFKL (Balitsky-Fadin-Kuraev-Lipatov)~\cite{bfkl}
evolution equation. 
Hence a pertinent and exciting question is whether these 
 $\ln 1/x$ contributions 
to the parton evolution can be observed experimentally.

Differences between different dynamical assumptions for the parton cascade
are expected to be most prominent in the 
phase space region towards the proton remnant
direction, i.e. away from the scattered quark. Here we investigate the 
region 
of central
 rapidity in the CMS system of the virtual photon
and the proton. In the HERA laboratory frame
this corresponds to a region of small polar angles and 
has been generically termed ``forward region"\footnote{
H1 uses a right-handed coordinate system with the $z$-axis defined
by the incident proton beam and the $y$-axis pointing upward. }.
In previous H1 analyses results have been
 presented on forward jet and forward inclusive
charged and neutral pion production~\cite{forwjets,h1fwdjet}, based
on data collected in 1994. A  measurement on forward jet 
production has been presented by
the ZEUS collaboration~\cite{zeusfwdjet}.
In this paper we
study forward single {\piz} production for a considerably larger data sample
than that of~\cite{h1fwdjet}, collected in 1996 and which allows 
the selection of particles with 
larger transverse momentum, $p_T$.
The production
 of high {\pt} particles is strongly correlated with  the emission of
 hard partons in QCD and is therefore sensitive to the 
dynamics of
 the strong interaction~\cite{kuhlenpt,h1pt,twthesis}.
An advantage of studying single particles, as opposed
to jets, is that no jet algorithm is needed 
and  the potential to reach smaller angles 
than is possible with jets with broad spatial extent. 
Furthermore, theoretical calculations at the parton level
can be convoluted with known fragmentation functions\cite{bkk},
allowing a direct comparison of the measurements  and theory.
The selection of high $p_T$ particles is also inspired by the 
proposal of Mueller~\cite{mueller} to select events where the 
photon virtuality $Q^2$ and transverse momentum 
squared of the parton emitted
in the parton cascade, $k_T^2$, are of similar magnitude, thereby 
suppressing the $k_T$ ordered DGLAP evolution with respect to 
the non-$k_T$ ordered BFKL evolution.

In this analysis {\piz}'s are selected 
in  DIS events at low $x$ in the 
region of momentum transfer  2 $<$ {\qsq} $<$ 70 GeV$^2$.
The {\piz}'s are required to have a polar angle 
in the lab frame between $5^\circ$ and 
$25^\circ$, and transverse momentum larger than  2.5~GeV in the
hadronic CMS (contrary to the analysis in ~\cite{h1fwdjet}, where 
a minimum transverse momentum  of 1 GeV was required in the laboratory frame).
Large transverse momenta in the hadronic CMS, as opposed to the laboratory 
system, are more directly related to hard 
subprocesses, since in the  quark parton model picture
the current quark has zero $p_T$ in the hadronic CMS.
The increased transverse momentum cut 
enhances the sensitivity to hard parton emission
in the QCD cascade and provides a hard scale for perturbative calculations.
It also reduces significantly the influence of
soft hadronization.

A  calculation based on pQCD
which uses the BFKL formalism for the perturbative part, and 
fragmentation functions for the hadronization,
is available~\cite{KwMaOu}
 and will be
compared with the data. In addition, 
models using ${\cal O}(\alpha_s)$ QCD 
matrix elements and parton cascades according to  DGLAP
evolution, and colour string hadronization, will be compared with the data.

\section{Experimental Apparatus}
\noindent
A detailed description of the H1 detector can be found 
elsewhere~\cite{h1nim}. The following section briefly describes the
components of the detector relevant for  this analysis.
 
The hadronic energy flow and the scattered electron are measured with a
liquid argon~(LAr) calorimeter and a backward SPACAL
calorimeter, respectively. The LAr calorimeter~\cite{larc}  extends
over the polar angle range $4^\circ < \theta <  154^\circ$ with full
azimuthal coverage. It consists of an electromagnetic section with
lead absorbers and a hadronic section with steel absorbers.
With about $44\,000$ cells in total, both sections are highly segmented 
in the transverse and 
the longitudinal direction, in particular in the forward region of the
detector. The total depth of
both sections varies between 4.5 and 8 interaction lengths in the
region $ 4^\circ < \theta < 128^\circ$. Test beam measurements of the
LAr~calorimeter modules showed an energy resolution 
of $\sigma_{E}/E\approx 0.50/\sqrt{E\;[{\rm GeV}]} \oplus 0.02$  for 
charged pions 
and of $\sigma_{E}/E\approx 0.12/\sqrt{E\;[{\rm GeV}]} \oplus 0.01$  for 
electrons~\mbox{\cite{larc}}. The hadronic energy measurement is
performed by applying a weighting technique in order to account for
the non-compensating nature of the calorimeter. The absolute scale of
the hadronic energy is presently known to $4\%$. The scale uncertainty
for electromagnetic energies is 3\%~\cite{elan} for the forward
region relevant for this analysis.
 
The SPACAL~\cite{spanim} is a lead/scintillating 
fibre calorimeter which covers the
region $ 153^\circ < \theta < 177.8^\circ$ with an
electromagnetic section
 and a hadronic section. The energy resolution for electrons
 is $7.5\,\%/\sqrt{E} \oplus
2.5\,\%$, the energy resolution for hadrons is 
$\sim 30\%$. The energy scale uncertainties
are 1$\%$ and 7$\%$ for the electrons and hadrons
respectively. The timing resolution of better than 1~ns in both
sections of the SPACAL is exploited to form a trigger decision and
reject background.

The calorimeters are surrounded by a superconducting solenoid
providing a uniform magnetic field of $1.15$ T parallel to the beam
axis in the tracking region. Charged particle tracks are measured in
the central tracker (CT) covering the polar angular range $ 25^\circ <
\theta < 155^\circ$ and  the forward tracking (FT) system,
covering the polar angular range $ 5^\circ < \theta < 25^\circ$. 
The CT consists of inner and outer cylindrical jet chambers, $z$-drift 
chambers and proportional chambers.  The jet chambers,  mounted
concentrically around the beam line,
 provide up to  65 space points in the radial plane
for tracks with sufficiently large transverse momentum. 
 
A backward drift chamber (BDC) in front of the SPACAL with an angular
acceptance of $151^\circ < \theta < 177.5^\circ$ serves to identify electron
candidates and to precisely measure their direction. Using information
from the BDC, the SPACAL and the reconstructed event vertex the 
polar angle of the scattered electron is known to about 0.7 mrad. 
 
The luminosity is measured using the reaction $ep\rightarrow ep\gamma$
with  two TlCl/TlBr crystal calorimeters installed in the HERA tunnel.
The electron tagger is located at $z=-33$ m and the photon tagger 
at $z=-103$ m from the interaction point in the direction of the outgoing
electron beam.

\section{Theoretical Predictions}

\noindent
Predictions for final state observables are available 
from Monte Carlo models using ${\cal O}(\alpha_s)$ matrix
elements and parton cascades according to the DGLAP evolution, 
and from numerical calculations
based upon the BFKL formalism. In the following we describe the 
models and calculations used.

\subsection{Phenomenological QCD Models}
Implementations of ${\cal O}(\alpha_s)$ matrix elements complemented by
parton showers based on the DGLAP splitting functions are available in
the programs LEPTO6.5 \cite{lepto} and HERWIG5.9 \cite{herwig}. 
The factorization and renormalization scales are
set to $Q^2$.  The
predictions of these models should  be valid in the 
region: 
${\alpha_{s}}(Q^{2})\, {\rm ln}(Q^{2}/Q_{0}^{2}) \sim 1$ and
${\alpha_{s}}(Q^{2})\, {\rm ln}(1/x) \ll 1$. 
In LEPTO the Lund string model as implemented in JETSET7.4 
\cite{jetset} is used to describe 
hadronization processes. LEPTO includes soft colour interactions
in the final state which can lead to events with a large rapidity gap.
HERWIG differs from LEPTO in that it also considers interference 
effects due to
colour coherence and uses the cluster fragmentation model for
hadronization. The versions of LEPTO6.5 and HERWIG5.9 
used consider only DIS processes in which the virtual
photon is  point-like.

Recently a model has been proposed (RAPGAP2.06 \cite{rapgap}) 
which is also based on the DGLAP formalism but includes
contributions from processes in which the virtual photon 
entering the scattering process can be resolved.
The
relative contribution from resolved photon processes depends on the scale at
which the virtual photon is probed. As in \cite{resstud} the
factorization and renormalization scale
in this paper is taken to be $Q^{2} + p_{T}^{2}$ ($p_{T}^{2}$ of the partons
from the hard subprocess).

The model calculations in this paper were made with the CTEQ4M~\cite{CTEQ4}
parton densities for the proton and the SAS-1D~\cite{sas} parton
densities for the virtual photon.
QED corrections are determined with the Monte Carlo program
DJANGO6.2~\cite{django}.
The contribution of photons emitted in the forward direction from
QED processes originating from the quarks in the proton, were found
to be negligible.%\cite{favart}.

In a previous paper~\cite{h1fwdjet} we  compared the results with 
ARIADNE~\cite{ariadne} and LDCMC~\cite{Lonnblad-ldc}.
 ARIADNE
provides an implementation of the Colour Dipole
Model (CDM) of a chain of independently radiating dipoles formed by
emitted gluons~\cite{cdm}. 
Unlike  LEPTO, the cascade of the CDM is not 
 ordered in transverse momentum.
For the present analysis it was confirmed that the predictions 
depend strongly on the parameters controlling the ``size" of the 
diquark and photon, and are therefore not explicitly compared with 
data in this paper\footnote{
A good description of the data presented in this paper 
can be achieved (using ARIADNE4.10) 
e.g. choosing PARA(10)=1.7 and PARA(14)=1.0~\cite{twthesis}.}.
 The linked dipole chain (LDC) model~\cite{ldc} is a reformulation of the 
CCFM~\cite{ccfm} equation, which forms a bridge between
the BFKL and DGLAP approaches. Calculations of the hadronic final 
state based on this approach are available with the LDCMC  
1.0
Monte Carlo which matches exact first order matrix elements with  
the LDC-prescribed initial and final state parton emissions. 
The model  however failed to describe the data in~\cite{h1fwdjet}
and is therefore not considered further in this paper.

\subsection{BFKL Calculation}

Recently {\piz} cross-sections have 
been calculated~\cite{KwMaOu} based on a modified BFKL evolution equation
in order ${\cal O}(\alpha_s)$ convoluted with {\piz} fragmentation
functions. The modified evolution equations include the so called 
``consistency constraint''~\cite{martinglx,martinf2}
which limits the gluon emission at each vertex in the cascade to 
the kinematically allowed region. It is argued that this constraint
embodies a major part of  the non-leading ${\rm ln}(1/x)$
contributions to the BFKL equation, which have been found to be very 
important~\cite{newbfkl}. The predictions of this modified BFKL
equation are therefore expected to be more reliable than those without
this constraint.
 The
parton densities and fragmentation functions used in the calculation
are taken from \cite{mrst} and \cite{bkk} respectively.
In this paper we compare ``set (iii)'' of~\cite{KwMaOu} to the data. In
this set 
the scale for the strong coupling constant 
$\alpha_S$ is taken to be the transverse 
momentum squared of the emitted partons, $k_T^2$, and the infrared
cut-off in the modified BFKL equation is taken to be 0.5 GeV$^2$.
Calculations with these parameters give a fair description of the 
forward jet cross-sections from~\cite{h1fwdjet} when taking into 
account hadronization corrections.
The predictions are labelled ``mod LO BFKL'' in the figures.

\section{Measurement}
\noindent
\subsection{Data Selection}
The analysis is based on data representing an integrated luminosity of
${\lum} = 5.8~{\rm pb}^{-1}$ taken by H1 during  1996.
Deep-inelastic scattering events are selected and the event kinematics
are calculated from the polar angle and the energy of
the scattered positron. 
The  
four momentum transfer squared, $Q^2$, and the inelasticity,
$y$, are related to these quantities (neglecting the
positron mass) by 
$Q^2 \,=\, 4\, E_e\, E_l\, {\rm cos}^{2}\,\frac{\theta_e}{2} \,$
and
$y \,=\, 1\,-\,\frac{E_e}{E_l}\, {\rm sin}^{2}\,\frac{\theta_e}{2}\,,$
where $E_l$ and $E_e$ are the energies of the incoming and the
scattered positron respectively, and $\theta_e$ is the polar angle of
the scattered positron.
Bjorken-$x$ is then given by
{\xbj}$\,=\,Q^{2}\,/\,(\,y\,{\cdot}\,s\,)$, where $s$ is the square of the
$ep$ center of mass energy.

Experimentally the scattered positron is defined to be the highest
energy cluster, i.e. localized energy deposit, in the SPACAL with a
cluster radius of less than 3.5~cm and an associated track in the BDC.
Experimental requirements based on the energy and the polar angle of the
scattered positron are used during the preselection but these are superseded
by  stronger kinematic cuts which
restrict the data to the range $0.1 < y < 0.6$ and 
$2 < Q^2 < 70$~GeV$^2$. The restricted $y$-range ensures that the particles
from the current  quark are detected in the central detector, and not 
in the forward region, and that the DIS kinematics can be well determined from
 the measurement of the scattered positron.
Photoproduction background is further reduced to a negligible level by
requiring $35 < \sum_j \, (E_{j}\,-\,p_{z,j}) < 70~{\rm GeV}$
\cite{f2pap} with $E_{j}$ and $p_{z,j}$ the energy and longitudinal
momentum of a particle respectively, and
where the sum extends over all detected particles in the
event, except for those in the small angle electron and photon tagger.
The reconstructed primary event vertex must have a $z$
coordinate  not more than 35~cm away from the nominal
interaction point.
The  trigger is based on  energy
depositions in the SPACAL and  
demands multiple track activity in the central tracker. For the
events used in this analysis the efficiency of this trigger is around
80$\%$, determined using data from an independent second 
trigger. 

After the selection about 600K events are
available for further analysis.

\subsection{Forward \boldmath{\piz}-Meson \unboldmath Selection}

A measurement of particle production at mid-rapidity in the hadronic
CMS system requires small
forward angles in the lab system. It is difficult to identify
individual charged particles in the forward direction in an environment with a
high density of charged particles. However the finely segmented H1 LAr
calorimeter  allows the measurement of {\piz}'s down to very small
angles. They 
are measured using the dominant decay channel 
{\piz} $\rightarrow 2\gamma$. The {\piz} candidates are selected in
the region  $5^{\circ} < {\thpi} < 25^{\circ}$, where {\thpi} is the polar
angle of the produced {\piz}. Candidates are required to have an
energy such that 
${\xpi }={\epi}/{\eprot} >$ 0.01, with {\eprot} the proton beam energy
(820 GeV), and 
a transverse momentum in the hadronic CMS, {\ptpi}, greater than  2.5~GeV.
At the high {\piz} energies considered here, the two photons from the
decay  cannot be separated, but appear as one object (cluster) in the
calorimetric response. Therefore, the standard method 
to identify {\piz}-mesons by reconstructing the
invariant mass from the separate measurement of the two decay photons
 is  not applicable.

In this paper, a detailed analysis of the longitudinal and transverse shape
of the energy depositions is performed
to separate electromagnetic from hadronic showers. 
This approach is based on the
compact nature of electromagnetic showers as opposed to showers of
hadronic origin, which are broader. 
The analysis of shower profiles is made possible by the fine
granularity of the calorimeter in the forward direction. It
has  a typical lateral
cell size of $3.5 \times 3.5~{\rm cm}^2$. This can be compared to the mean
Moliere radius $\bar{R}_m$ which is  3.6~cm and the mean radiation length
$\bar{X}_0$ which is  1.6 cm. The calorimeter has a 
four-fold longitudinal 
segmentation for the electromagnetic section which has a thickness of
20 to 25 radiation lengths $X_0$.
The main experimental challenge in this analysis is the
high activity in this region of phase space, with hadronic showers
``masking'' the clear electromagnetic signature
from the {\piz} $\rightarrow 2\gamma$ decay. 
The overlap of a {\piz} induced cluster
with another hadron is mainly responsible for losses
of {\piz} detection efficiency, since the distortion of the shower shape
estimators it causes will, in many cases, lead to the rejection of the
cluster candidate.

The reconstruction of LAr data is optimized to contain all  the energy
of an electromagnetic shower in one cluster~\cite{h1nim}.
A {\piz}-meson candidate is  required to be a cluster 
with more than 90$~\%$ of
its  energy deposited in the electromagnetic part of the LAr
calorimeter.  A ``hot'' core consisting of the most energetic
group of contiguous electromagnetic calorimeter cells of a cluster, 
which must include the hottest cell, is defined for each
candidate~\cite{epsep}. More than 50~$\%$ of the cluster energy is
required to be deposited in this core. The lateral spread of the shower
is quantified in terms of lateral shower moments calculated relative to 
the shower's principal axis~\cite{epsep} and required to be 
less than 4 cm~\cite{twthesis}. The longitudinal shower shape is used
as a selection criterion via the fraction of the shower's energy
deposited in each layer of cells in the electromagnetic  part
of the calorimeter. The precise specifications of these layers can be
found in~\cite{epsep}. 
The part of the cluster's energy measured in the 
second layer minus that  measured in the fourth layer 
 is required to be more than 40\% of the total cluster energy. This
 selects showers which start to develop close to the
calorimeter surface and are well contained in the electromagnetic part
of the calorimeter, as expected for showers of electromagnetic origin.
As mentioned above, with this selection one cannot distinguish
photons from a {\piz}-meson decay and photons from other
sources. The high  energy required in the selection, however,
ensures that contributions from sources other than high energy
{\piz}-mesons (such as prompt photon production) are at a negligible
level~\cite{klm}. The influence of $\eta$-meson production is corrected
for in the analysis. Uncertainties in  the relative
$\eta$ and {\piz}  production rates in the Monte Carlo models used
have been studied and were found to have an negligible effect on the results.

With this selection about 1700 (600) {\piz} candidates are found  in the
kinematic range $5^{\circ} < {\thpi} < 25^{\circ}$, ${\xpi} > 0.01$
and ${\ptpi} >$ 2.5 (3.5)~GeV, with a detection  efficiency 
better than 45$\%$.
Monte Carlo studies, using a detailed simulation of the H1 detector
for a sample of DIS events,
yield a purity of about 70\% for the selected {\piz}-meson sample.
The impurities are due to misidentified hadrons and from
secondary interactions of charged hadrons with 
passive material in the detector (between one and two radiation 
lengths in the
forward region). 
These studies show that 
 less than 10$\%$  of the selected {\piz} candidates 
stem from  secondary scattering of
charged hadrons with passive material in the forward region, where the
amount of material between the interaction point and the calorimeter
surface is largest. 

The determination of the {\piz} acceptance and purity depends
only on the particle density and energies in the forward calorimeter.
To ensure that the different Monte Carlo
 models used are in reasonable agreement with
the data in this respect the transverse energy flow, $E_T$, around the {\piz}
candidate clusters was studied in detail.
%, similar to a jet profile. 
Both the $E_T$ flow and the
$E_T$ spectra in the tail of the $E_T$ flow distributions are
reasonably well
described by the models used to determine the detector
corrections~\cite{twthesis}. 
Of the two models used, however, ARIADNE showed a higher particle
density while LEPTO has a lower
particle density than the data.
Remaining differences of the detector corrections
determined with the two Monte Carlo models are therefore used to estimate the
systematic error. When the energy and transverse
momentum requirements are lowered, the 
two photons from the {\piz} decay become separable and a clear {\piz} mass
peak can be observed which is also well reproduced by the H1 detector
simulation. The same method  of selecting
{\piz}-mesons as outlined above was used in a 
previous H1 analysis~\cite{h1fwdjet}, where a measurement of charged
particles in the same region was also performed. 
The measured {\piz} cross-sections were found to agree 
well with the average of the $ \pi^{+} $ and $\pi^{-}$ cross
sections. Furthermore, the results of the
analysis of the 1994 data are found to be in good agreement with the
present analysis in their overlapping phase space regions.

\section{Results}
\noindent
The  experimental
results of the analysis are presented as differential $ep$ cross-sections of
forward {\piz}-meson production as a function of $Q^2$, and 
as a function of {\xbj},
{\etapi} and {\ptpi} in three regions of {\qsq} for {\ptpi} 
$> 2.5$~GeV. The pseudorapidity {\etapi} is given by 
$-\ln\left[ {\rm tan}\left(\theta/2\right)\right]$ with $\theta$ being the
polar angle of the {\piz} in the  
laboratory frame.
 In addition the {\piz} cross-sections as a function of {\xbj}
and {\qsq} are measured for data with the  threshold of the
{\piz} transverse momentum  increased to 
{\ptpi}~$>$ 3.5~GeV. 
An increased {\ptpi} threshold is expected to enhance the
sensitivity to hard parton emission in the parton cascade. 
The phase space is given by 0.1 $<$ {\ybj} $<$
0.6, 2 $<$ {\qsq} $<$ 70 GeV$^2$, $5^\circ <$ {\thpi} 
$< 25^\circ$ and {\xpi} $=$ {\epi}$/${\eprot} $>$ 0.01, in addition to
the {\ptpi} thresholds given above. {\thpi}, {\epi} and {\eprot}   are 
measured in
the H1 laboratory frame; {\ptpi} is calculated in the hadronic CMS.
The measurement extends down to  
{\xbj} $=$ 4$\cdot$10$^{-5}$, covering two orders of magnitude in 
Bjorken-$x$.

All observables are corrected for detector effects and for the
influence of QED radiation by a bin-by-bin unfolding procedure.
The detector effects include the efficiency, purity and acceptance of the
{\piz}-meson identification as well as contributions from secondary
scattering in passive material. The correction functions are obtained
with two different models (ARIADNE and LEPTO) and 
detailed detector simulation. The final correction is
performed with the average of the two models. 
The remaining background
from photoproduction in the data sample has been studied using a sample of
photoproduction Monte Carlo events (PHOJET~\cite{phojet}) representing
an integrated luminosity of about 1~pb$^{-1}$. The contribution from
such events is found to be negligible in all bins.

The typical total systematic uncertainty is 15-25$\%$, compared to a
statistical uncertainty of about 10$\%$. Contributions to the
systematic error include: the uncertainty of the luminosity
measurement (1.8$\%$), the statistical uncertainty in the
determination of the trigger efficiency (5$\%$), the uncertainty of
the electromagnetic energy scale of the LAr (3\%) and the SPACAL (1\%)  
 calorimeters which each contribute 5-10$\%$, the variation of {\piz}-meson
selection and acceptance requirements within the resolution of the 
reconstructed quantities
(5-10$\%$), and the model dependence of the 
bin-by-bin correction procedure using differences between
ARIADNE and LEPTO
 (5-10$\%$). 

The cross-sections as a function of {\xbj}, shown in Fig.~\ref{fig:xbja}(a), 
exhibit a strong rise towards small {\xbj}. 
In this and the following figures the inner error bars give the 
statistical errors, while the outer error bars give the 
statistical and systematical error added quadratically.
It is of interest to note that 
 the rise in $x$ in Fig.~\ref{fig:xbja}(a) is similar
to  the  rise of the total inclusive cross-section as measured
e.g. in~\cite{h1f2jer}. This is demonstrated in Fig.~\ref{fig:xbja}(b),
which shows the rate  of {\piz}-meson production in DIS as a function of
{\xbj} obtained by dividing the cross-section shown in
Fig.~\ref{fig:xbja}(a) by the
inclusive $ep$ cross-section in each bin of {\xbj} and {\qsq}.
The inclusive cross-section is calculated by integrating the H1
QCD fit to the 1996 structure function data as presented
in~\cite{h1f2jer} for every bin  of
inclusive {\piz}-meson cross-sections. Note, however, that
the {\piz} rate increases  with increasing $Q^2$.
The $x$-independence seen in Fig.~\ref{fig:xbja}(b) in 
a fixed $Q^2$ interval implies that the {\piz} rate for particles
  with a $p_T$ above the cut-off
and within the selected kinematical region, 
is independent of $W$, the  hadronic invariant mass of the 
photon-proton system.

The shapes of
$d{\sigma_{\pi}}/d{\etapi}$ and $d{\sigma_{\pi}}/d{\ptpi}$ 
(Fig.~\ref{fig:etapt}) show no significant dependence
on  {\qsq}.
 The measurements of the latter extend to values of
transverse momenta as high as 8~GeV. Since {\etapi} is 
measured in the laboratory
frame, mid-rapidity in the hadronic CMS corresponds approximately to
{\etapi}$= 2$ in Figure \ref{fig:etapt} (a).
Figure
\ref{fig:q2a} shows the inclusive {\piz}-meson cross-section as a
function of {\qsq}. The cross-section falls steeply with increasing $Q^2$.
Figure \ref{fig:xbjq2b} finally shows $d{\sigma_{\pi}}/d${\qsq} and
$d{\sigma_{\pi}}/d${\xbj} for the higher threshold of {\ptpi}~$>$
3.5~GeV. No significant change in shape of the distributions occurs
 when the {\ptpi} threshold is raised, but the 
 cross-sections are reduced by about a factor of three.

All differential cross-sections are compared to three predictions
based  on different QCD approximations.
The DGLAP prediction for pointlike virtual photon scattering
(including parton showers)
as given by LEPTO6.5 falls
clearly below the data. There is still a fair agreement with data
for the highest $x$ and $Q^2$ bins, shown in Fig.~\ref{fig:xbja}(a)
and Fig.~\ref{fig:q2a}, but differences occur in the low-$x$
region. These 
are as large as a factor of five in the lowest $x$ region.
LEPTO also fails to describe the ratio in Fig.~\ref{fig:xbja}(b),
and shows a strong decrease with decreasing $x$.
The mechanism of emitting partons according to the DGLAP
splitting functions, combined with pointlike virtual 
photon scattering only, is clearly not supported by
the data, in particular at low $x$. 
The LEPTO prediction is based on about seven times the integrated
luminosity of the data.
Comparisons to HERWIG5.9 (not shown)
lead to similar conclusions~\cite{twthesis}.

A considerable improvement of the description of the data
is achieved by a model which considers  additional  processes where
the virtual photon entering the scattering process is resolved. 
This approach can be regarded as  an effective resummation
of higher order corrections. It also provides a smooth transition towards
 the limit of $Q^2 =0$, i.e. 
photoproduction, where resolved processes dominate in the  HERA regime.
Such a
prediction is provided by RAPGAP2.06~\cite{rapgap}. 
In Fig.~\ref{fig:xbja}(a) RAPGAP2.06 
predicts a  cross-sections very close to the measured
distributions 
with the exception of the  lowest  $Q^2$ bin where the prediction is 
too low.
 All predicted cross-sections increase by up
to 30$\%$ when the scale in the hard scattering is increased 
from $Q^{2} + p_{T}^{2}$
 to $Q^{2} + 4p_{T}^{2}$~\cite{twthesis}, and therefore do not improve
 the overall 
description significantly. 
Hence RAPGAP, with the parton distributions used here, 
does not describe the low-$x$ behaviour of the data over the full range.
The RAPGAP prediction is
 based on approximately four  times the integrated luminosity of the data.

The ARIADNE model (not shown), with parameters as given before,
can describe the data presented in this paper~\cite{twthesis}, but it remains
to be shown whether this choice allows for a consistent description
of other aspects of the DIS final state data. Moreover, a moderate variation 
of these parameters leads to large changes in the prediction.

Next we compare the data with a 
   prediction of the {\piz}-meson cross-section based on a 
 modified LO BFKL parton calculation convoluted with {\piz} fragmentation
functions.
The predictions obtained with these calculations turn out to be in good
agreement with the neutral pion cross-sections  measured in most of the
available phase space, but are below the data at
the lowest values of $Q^2$.
The calculation also describes well the ratio shown in Fig.~\ref{fig:xbja}(b),
except possibly at the largest $Q^2,x$ bin. This ratio has been calculated
by using the corresponding prediction for the inclusive cross section, 
based on the BFKL formalism~\cite{martinf2}.

The BFKL predictions involve a cut-off parameter in the transverse momentum
squared $k^2_T$ of the partons: 
$k_{0}^2$=0.5 GeV$^2$ and a
 choice of renormalization
scale. It is shown in~\cite{KwMaOu}  that a variation of 
$k_0^2$  by a 
 factor of two leads to less than a  10\% change of the cross-sections.
The scale dependence 
is larger; a change from $k_T^2$ to $k_T^2/4$ leads to an 
approximate increase of 
 60\% of the cross-sections. 
This affects  mostly the normalization, but not the shape of the 
distributions.

 The good agreement between this
 prediction and the data suggests that the modified 
 BFKL evolution equation, using the 
  consistency constraints, is a good
 approximation for low-$x$ evolution in the considered phase space.
 This, in turn, can then be interpreted as a sign of the  
 experimental manifestation of  leading $\ln 1/x$ terms which are 
 anticipated in  pQCD evolution. 

\section{Conclusions}
\noindent

Differential cross-sections of forward {\piz} production have been measured 
for particles with $\ptpi > $ 
2.5 (3.5) GeV, $5^\circ <$ {\thpi}$< 25^\circ$ and {\xpi} $=$ {\epi}$/${\eprot} $>$ 0.01, for DIS events with
$0.1<{\ybj}<$
0.6 and  2 $<$ {\qsq} $<$ 70 GeV$^2$.
The  data are  sensitive to  QCD parton dynamics
at low $x$ (high parton density) and mid-rapidity in the hadronic
CMS system. 
They discriminate between different approximations to QCD evolution
            in the new regime opened
           up by HERA.
The data show a strong rise of the cross section with decreasing $x$.
This rise is  similar to the rise of the inclusive cross section.

Models using ${\cal O}(\alpha_s)$
QCD matrix elements and parton cascades according to the DGLAP
splitting functions cannot describe the differential neutral pion
cross-sections at low $x$. Inclusion of processes in which the virtual
photon is resolved improves the agreement with the data, but 
does not provide a satisfactory description in the full $x$ and $Q^2$ range.
A calculation based on the  BFKL formalism is in good agreement with the
data, particularly for the shape description, 
but the absolute normalization remains
strongly affected by the scale uncertainty.
So far the data in the phase space selected in this analysis could not be
confronted with a next-to-leading (NLO) order 
 prediction -- either for DGLAP or for  BFKL. 
More definite conclusions  therefore have to be delayed until such
calculations become available.

\section*{Acknowledgments}
\noindent
We wish to thank A.D. Martin, J.J. Outhwaite and A.M.~Stasto for useful
discussions.
We are grateful to the HERA machine group whose outstanding efforts have made
and continue to make this experiment possible. We thank the engineers and
technicians for their work constructing and maintaining the H1 detector, our
funding agencies for financial support, the DESY technical staff for continual
assistance and the DESY directorate for the hospitality which they extend to the
 non-DESY members of the collaboration.

\clearpage

\bibliography{pi096}

\clearpage

\begin{table}[t]
\begin{center}
\begin{small}
\renewcommand{\arraystretch}{1.25}
% x,eta,pt table
\begin{tabular}{|c|c||c|c||c|c|}
\hline
\multicolumn{1}{|c|}{$x \cdot 10^{4}$}&
 \multicolumn{1}{|c||}{${(\frac{{\rm d}
{\sigma_{\pi}}}{{\rm d}x})}^{\pm stat}_{\pm tot}$ /nb} &
\multicolumn{1}{|c|}{$\eta_{\pi}$}&
 \multicolumn{1}{|c||}{${(\frac{{\rm d}
{\sigma_{\pi}}}{{\rm d}\eta_{\pi}})}^{\pm stat}_{\pm tot}$ /pb} &
\multicolumn{1}{|c|}{$p_{T,\pi}^{\star}$ /GeV}&
 \multicolumn{1}{|c|}{${(\frac{{\rm d}
{\sigma_{\pi}}}{{\rm d}p_{T,\pi}^{\star}})}^{\pm stat}_{\pm tot}$ /$\frac{\rm pb}{\rm GeV}$}\\

\hline
\hline
\multicolumn{6}{|c|}{$2.0\,<\,Q^{2}\,<\,4.5~{\rm GeV}^{2}$ \hspace{2cm} 
$p_{T,\pi}^{\star}\,>\,2.5$~GeV (hcms)} \\
\hline
0.42 -- 0.79 & 1189$^{\pm 134}_{\pm 188}$ &
1.50 -- 1.93 & 141$^{\pm 16}_{\pm 28}$    & 
2.50 -- 2.80 & 187$^{\pm 20}_{\pm 92}$    \\

0.79 -- 1.1  & 1261$^{\pm 143}_{\pm 253}$ &
1.93 -- 2.23 & 165$^{\pm 20}_{\pm 31}$    &
2.80 -- 3.30 & 132$^{\pm 13}_{\pm 22}$    \\

1.1  -- 1.7  & 970$^{\pm 96}_{\pm 201}$   &
2.23 -- 2.56 & 129$^{\pm 15}_{\pm 29}$    &
3.30 -- 4.00 &  65.3$^{\pm 7.5}_{\pm 17.5}$    \\

1.7  -- 2.5  & 609$^{\pm 75}_{\pm 121}$   &
2.56 -- 2.85 & 108$^{\pm 13}_{\pm 26}$    &
4.00 -- 5.20 &  25.1$^{\pm 3.4}_{\pm 4.2}$     \\

2.5  -- 4.2  & 160$^{\pm 27}_{\pm 52}$    &
2.85 -- 3.13 & 111$^{\pm 15}_{\pm 26}$    &
5.20 -- 8.00 &   6.48$^{\pm 1.23}_{\pm 1.86}$    \\
\hline
\hline
\multicolumn{6}{|c|}{$4.5\,<\,Q^{2}\,<\,15.0~{\rm GeV}^{2}$ \hspace{2cm}
$p_{T,\pi}^{\star}\,>\,2.5$~GeV (hcms)} \\
\hline
1.1  -- 2.0  & 398$^{\pm 43}_{\pm 61}$    &
1.50 -- 1.91 & 103$^{\pm 11}_{\pm 16}$    &
2.50 -- 2.80 & 157$^{\pm 15}_{\pm 74}$     \\

2.0  -- 2.9  & 366$^{\pm 41}_{\pm 59}$    & 
1.91 -- 2.22 & 145$^{\pm 15}_{\pm 29}$    &
2.80 -- 3.40 &  93.5$^{\pm 7.8}_{\pm 21.7}$    \\

2.9  -- 3.9  & 337$^{\pm 39}_{\pm 64}$    &
2.22 -- 2.50 & 131$^{\pm 14}_{\pm 21}$    &
3.40 -- 4.10 &  56.9$^{\pm 6.4}_{\pm 9.5}$     \\

3.9  -- 5.5  & 193$^{\pm 21}_{\pm 34}$    &
2.50 -- 2.82 & 115$^{\pm 12}_{\pm 21}$    &
4.10 -- 5.20 &  16.6$^{\pm 2.4}_{\pm 3.8}$     \\

5.5  -- 11   &  63.5$^{\pm 6.8}_{\pm 13.8}$ &
2.82 -- 3.13 & 89.0$^{\pm 10.2}_{\pm 22.1}$ &
5.20 -- 8.00 &   4.45$^{\pm 0.75}_{\pm 1.18}$    \\
\hline
\hline
\multicolumn{6}{|c|}{$15.0\,<\,Q^{2}\,<\,70.0~{\rm GeV}^{2}$ \hspace{2cm} 
$p_{T,\pi}^{\star}\,>\,2.5$~GeV (hcms)} \\
\hline
3.9  -- 7.9  & 72.4$^{\pm 8.7}_{\pm 16.8}$    &
1.50 -- 1.93 & 66.2$^{\pm  7.9}_{\pm 14.9}$   &
2.50 -- 2.90 & 89.0$^{\pm 9.5}_{\pm 34.7}$    \\

7.9  -- 13   & 55.8$^{\pm 6.3}_{\pm 12.9}$    &
1.93 -- 2.26 & 93.5$^{\pm 11.2}_{\pm 17.5}$   &
2.90 -- 3.50 & 54.0$^{\pm 5.9}_{\pm 8.4}$     \\

13   -- 19   & 43.1$^{\pm 5.5}_{\pm 9.8}$     &
2.26 -- 2.69 & 73.2$^{\pm  8.1}_{\pm 13.5}$   &
3.50 -- 4.70 & 19.9$^{\pm 2.5}_{\pm 3.9}$     \\

19   -- 63   &  6.44$^{\pm 0.74}_{\pm 1.23}$  &
2.69 -- 3.13 & 43.6$^{\pm  5.4}_{\pm 10.6}$   &
4.70 -- 8.00 &  4.72$^{\pm 0.72}_{\pm 0.94}$    \\

\hline                                     
\end{tabular}
\vspace{2mm}

% Q2 table
\begin{tabular}{|c|c||c|c|}
\hline
\multicolumn{1}{|c|}{$Q^2$ /GeV$^2$}&
\multicolumn{1}{|c||}{${\left({{\rm d}
        {\sigma_{\pi}}}/{{\rm d}Q^{2}}\right)}^{\pm stat}_{\pm tot}$ 
  /$\frac{\rm pb}{\rm GeV^2}$} &
\multicolumn{1}{|c|}{$Q^2$ /GeV$^2$}&
\multicolumn{1}{|c|}{${\left({{\rm d}
        {\sigma_{\pi}}}/{{\rm d}Q^{2}}\right)}^{\pm stat}_{\pm tot}$ 
  /$\frac{\rm pb}{\rm GeV^2}$} \\
\hline
\hline
\multicolumn{4}{|c|}{$2.0\,<\,Q^{2}\,<\,70.0~{\rm GeV}^{2}$ \hspace{2cm} 
$p_{T,\pi}^{\star}\,>\,2.5$~GeV (hcms)} \\
\hline
 2.00 --  2.40 & 139$^{\pm      16}_{\pm      29}$    &
 6.50 --  9.10 &  19.0$^{\pm     1.8}_{\pm     3.6}$   \\
 2.40 --  3.00 & 106$^{\pm      11}_{\pm      23}$    &
 9.10 -- 13.0  &  11.5$^{\pm     1.1}_{\pm     1.8}$   \\
 3.00 --  3.80 &  78.2$^{\pm     7.3}_{\pm    13.3}$  &
13.0  -- 20.0  &   7.12$^{\pm    0.68}_{\pm    1.23}$  \\
 3.80 --  4.90 &  54.5$^{\pm     4.8}_{\pm    10.5}$  &
20.0  -- 32.0  &   3.64$^{\pm    0.36}_{\pm    0.63}$  \\
 4.90 --  6.50 &  35.5$^{\pm     3.1}_{\pm     5.3}$  &
32.0  -- 70.0  &   1.08$^{\pm    0.10}_{\pm    0.22}$  \\
\hline                                     
\end{tabular}
\vspace{2mm}

% Q2,x table for higher pt cut
\begin{tabular}{|c|c||c|c|}
\hline
\multicolumn{1}{|c|}{$x \cdot 10^{4}$}&
\multicolumn{1}{|c||}{${\left({{\rm d}
        {\sigma_{\pi}}}/{{\rm d}x}\right)}^{\pm stat}_{\pm tot}$ 
  /nb} &
\multicolumn{1}{|c|}{$Q^2$ /GeV$^2$}&
\multicolumn{1}{|c|}{${\left({{\rm d}
        {\sigma_{\pi}}}/{{\rm d}Q^{2}}\right)}^{\pm stat}_{\pm tot}$ 
  /$\frac{\rm pb}{\rm GeV^2}$} \\
\hline
\hline
\multicolumn{4}{|c|}{$2.0\,<\,Q^{2}\,<\,70.0~{\rm GeV}^{2}$ \hspace{2cm} 
$p_{T,\pi}^{\star}\,>\,3.5$~GeV (hcms)} \\
\hline
1.0  -- 1.9  & 374$^{\pm  41}_{\pm   70}$    &
2.00 -- 3.00 & 35.0$^{\pm 4.4}_{\pm  7.7}$   \\
1.9  -- 3.0  & 249$^{\pm  32}_{\pm   45}$    &
3.00 -- 4.90 & 22.5$^{\pm 2.3}_{\pm  3.9}$   \\
3.0  -- 5.0  & 165$^{\pm  21}_{\pm   30}$    &
4.90 -- 9.10 & 10.0$^{\pm 1.0}_{\pm  1.5}$   \\
5.0  -- 11   & 60.3$^{\pm 6.5}_{\pm  10.5}$  &
9.10 -- 20.0 & 3.20$^{\pm 0.35}_{\pm 0.55}$  \\
11   -- 63   & 5.90$^{\pm 0.69}_{\pm 1.50}$  &
20.0 -- 70.0 & 0.64$^{\pm 0.07}_{\pm 0.11}$  \\

\hline                                     
\end{tabular}
\end{small}
\end{center}

\caption{The numerical values of the inclusive differential {\piz}-meson cross
  sections as shown in Figures 1-4. 
  The value of the cross section is presented together with the respective 
  statistical and total error. The phase space is given by $0.1 < y < 0.6$, 
    $5^\circ < \theta_{\pi} < 25^\circ$ and $x_{\pi} > 0.01$ in
    addition to the restrictions given in the Table.}
\end{table}

\begin{figure}[t]
\begin{center}
\includegraphics{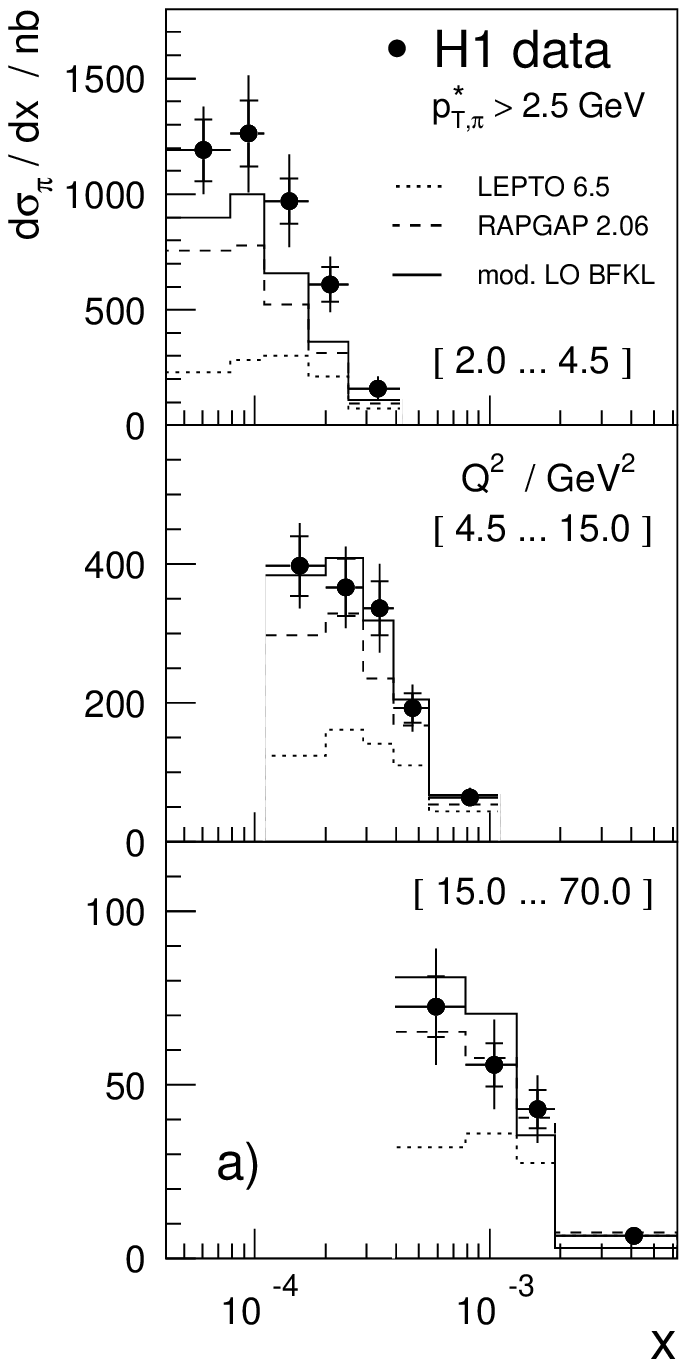}
\includegraphics{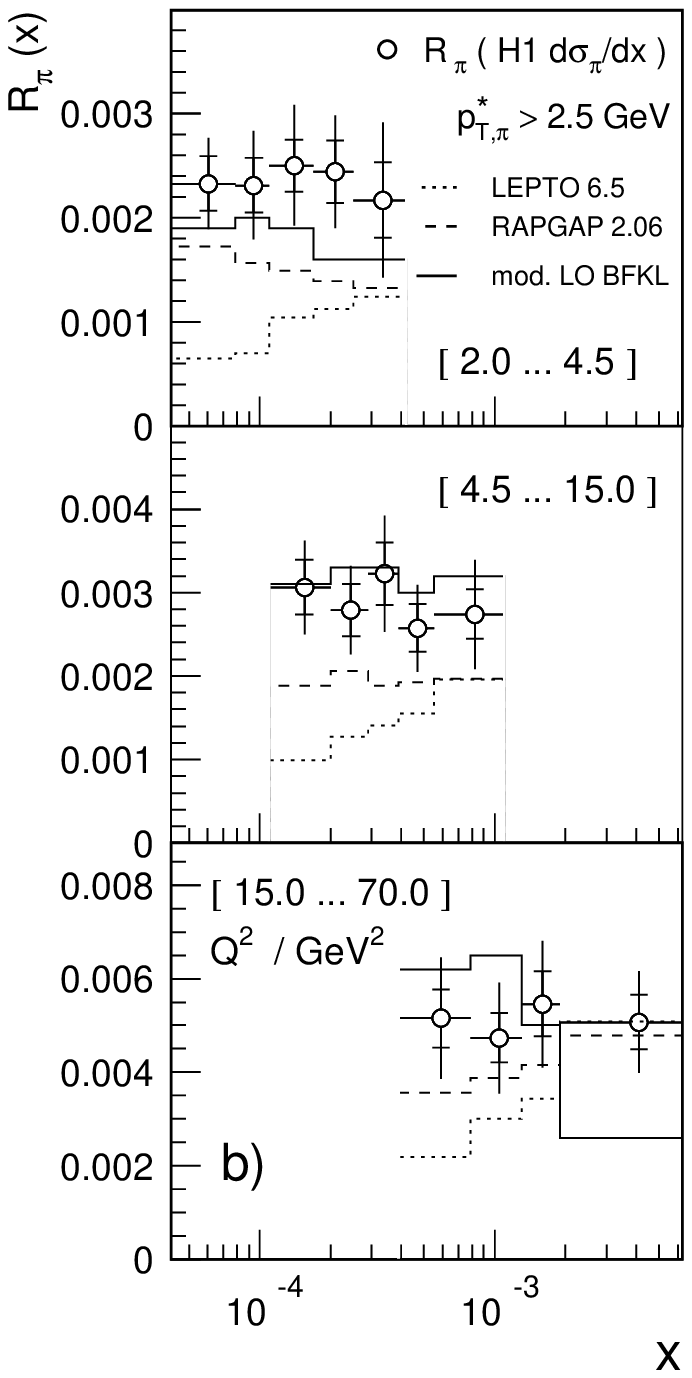}
\end{center}
\caption{Inclusive {\piz}-meson production cross-sections as a
  function of {\xbj} for {\ptpi} $> 2.5$~GeV in three
  regions of {\qsq} (a). The phase space is given by 0.1 $<$ {\ybj} $<$ 0.6,  
  $5^\circ <$ {\thpi} $< 25^\circ$ and 
  {\xpi} $=$ {\epi}$/${\eprot} $>$ 0.01 
  and the $Q^2$ ranges given in 
  the figure. 
  The inner error bars are 
statistical, and the outer error bars give the 
statistical and systematical error added quadratically.
  {\thpi} and {\xpi} are measured in the H1
  laboratory frame, {\ptpi} is calculated in the hadronic CMS. The
   QCD models RAPGAP (sum of direct and resolved contributions) and LEPTO
  are compared to the data. Also shown is the prediction of the modified LO
  BFKL calculation by Kwiecinski, Martin and Outhwaite. Figure (b)
  shows the rate  of {\piz}-meson production in DIS as a function of {\xbj}
  obtained by dividing the cross-section shown in  (a) by the
  inclusive $ep$ cross-section in each bin of {\xbj} and {\qsq}.}
\label{fig:xbja}
\end{figure}

\begin{figure}[t]
\begin{center}
\includegraphics{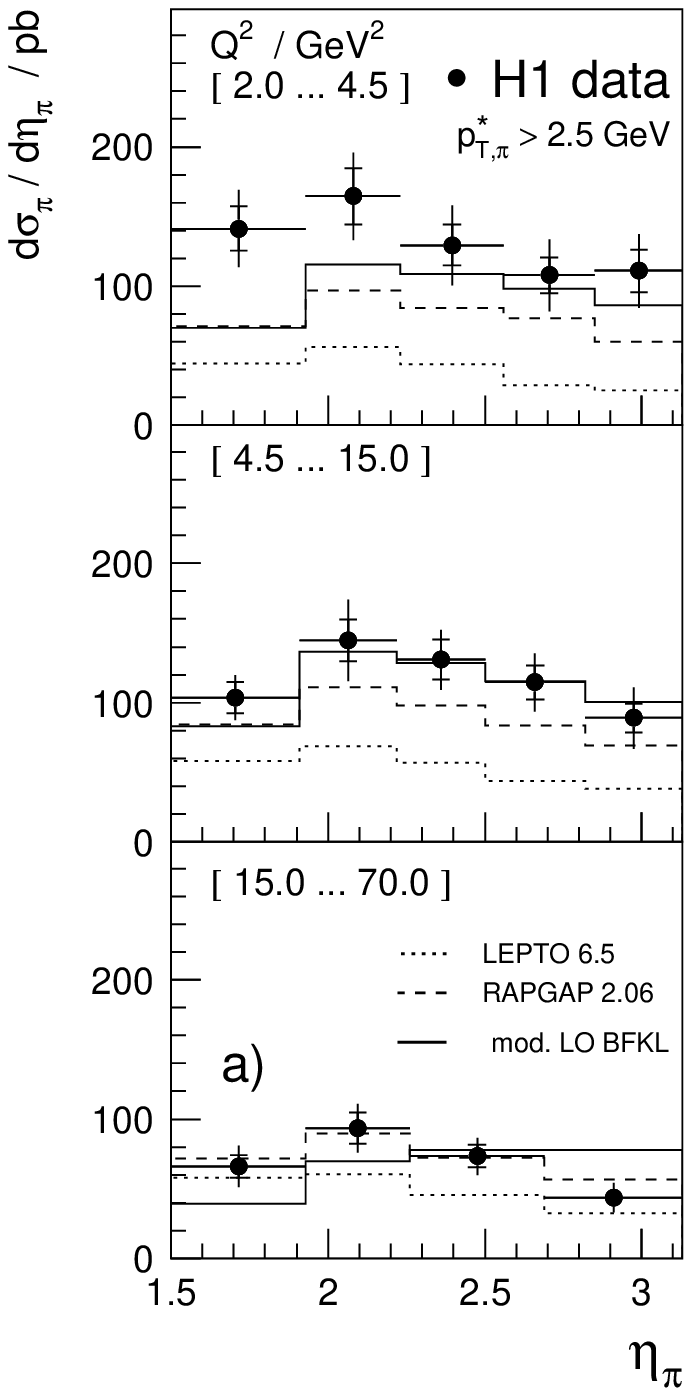}
\includegraphics{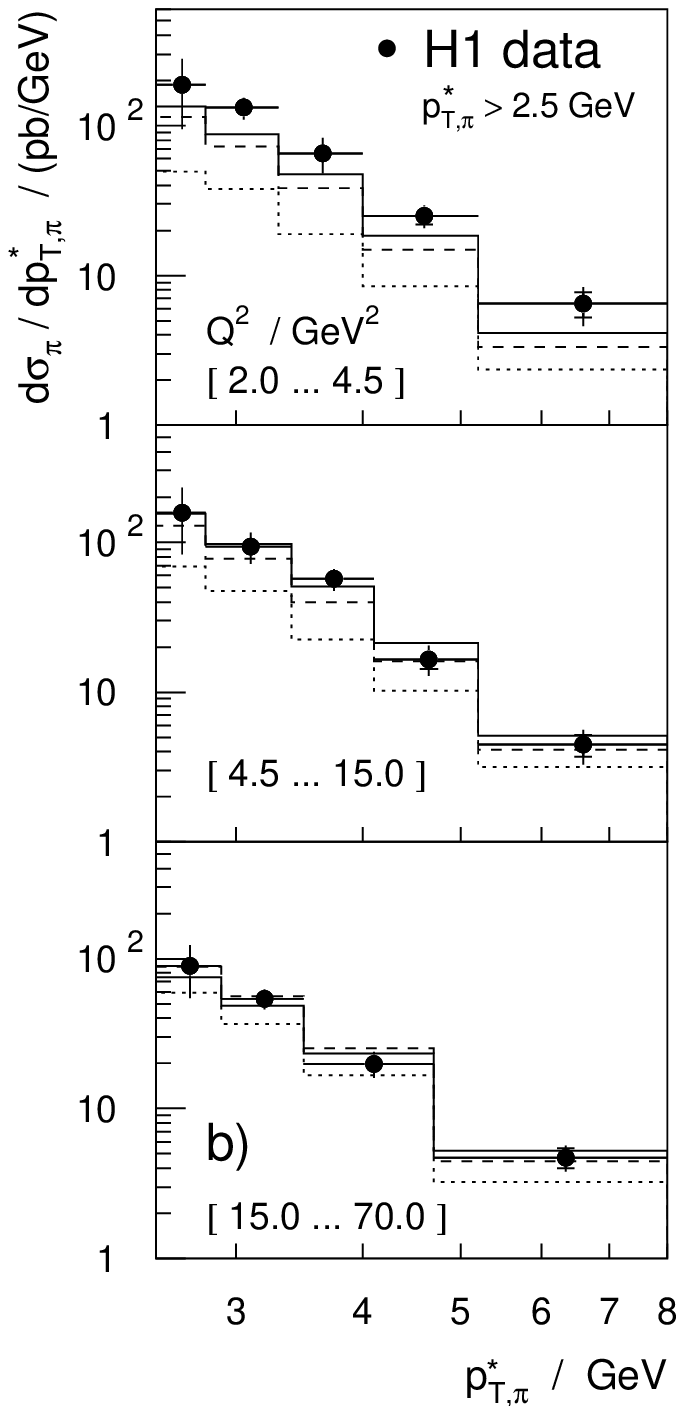}
\end{center}
\caption{Inclusive {\piz}-meson production cross-sections as a
  function of {\etapi} (a) and {\ptpi} (b) for {\ptpi} $> 2.5$~GeV in three
  regions of {\qsq}. The phase space is given by 0.1 $<$ {\ybj} $<$ 0.6,  
  $5^\circ <$ {\thpi} $< 25^\circ$ and 
  {\xpi} $=$ {\epi}$/${\eprot} $>$ 0.01
 and the $Q^2$ ranges given in  
  the figure. 
 The inner error bars are 
statistical, and the outer error bars give the 
statistical and systematical error added quadratically.
{\thpi} and {\xpi} are measured in the H1
  laboratory frame, {\ptpi} is calculated in the hadronic CMS. The
   QCD models RAPGAP (sum of direct and resolved contributions) and LEPTO
  are compared to the data. Also shown is the prediction of the
  modified LO BFKL calculation by Kwiecinski, Martin and Outhwaite.}
\label{fig:etapt}
\end{figure}

\begin{figure}[t]
\begin{center}
\includegraphics{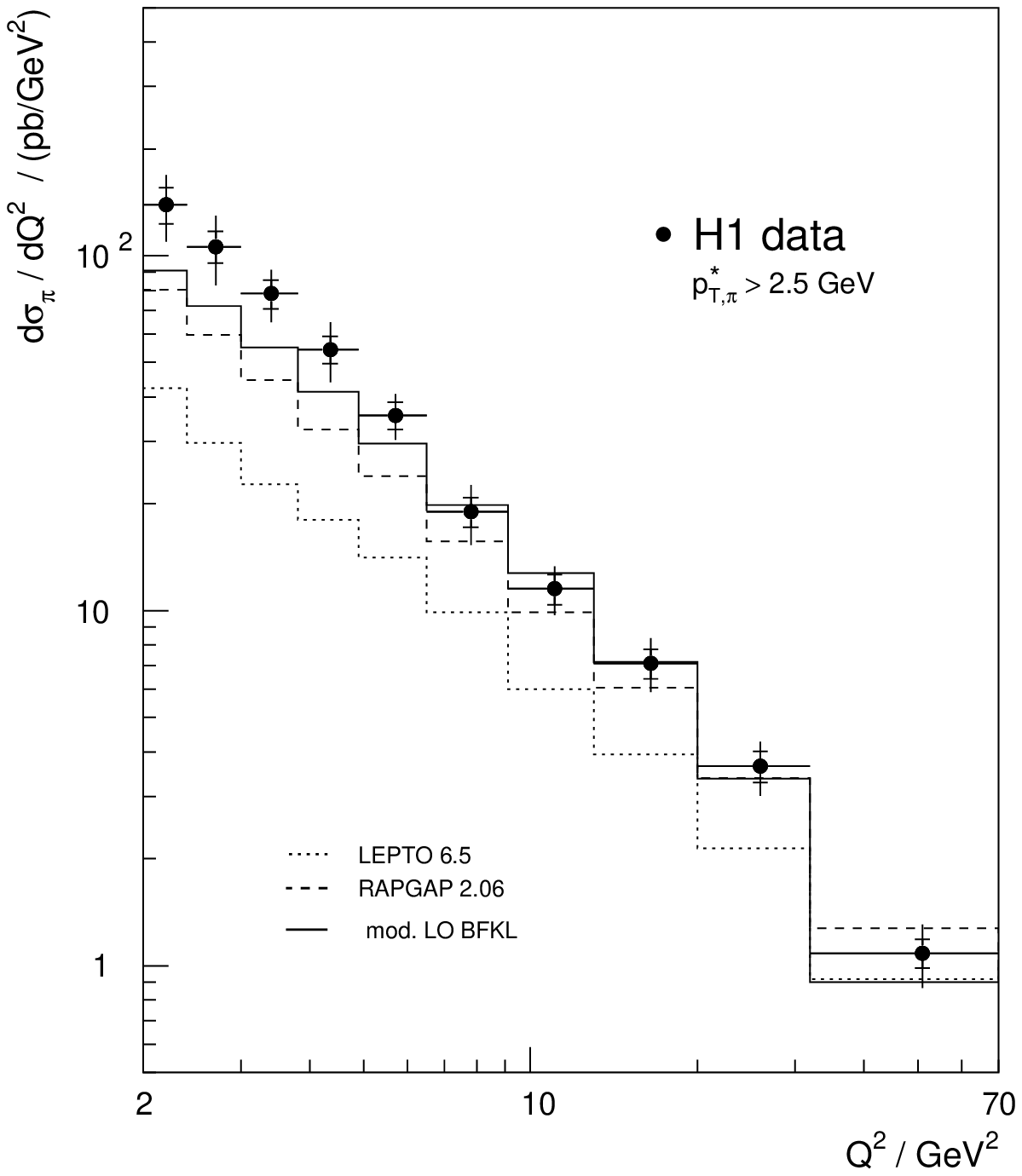}
\end{center}
\caption{Inclusive {\piz}-meson production cross-sections as a
  function of {\qsq} for {\ptpi} $> 2.5$~GeV. 
  The phase space is given by 0.1 $<$ {\ybj} $<$ 0.6,  
  $5^\circ <$ {\thpi} $< 25^\circ$ and 
  {\xpi} $=$ {\epi}$/${\eprot} $>$ 0.01.
 The inner error bars are 
statistical, and the outer error bars give the 
statistical and systematical error added quadratically.
 {\thpi} and {\xpi} are measured in the H1
  laboratory frame, {\ptpi} is calculated in the hadronic CMS. The
   QCD models RAPGAP (sum of direct and resolved contribution) and LEPTO
  are compared to the data. Also shown is the prediction of the
  modified LO BFKL calculation by Kwiecinski, Martin and Outhwaite.}
\label{fig:q2a}
\end{figure}

\begin{figure}[t]
\begin{center}
\includegraphics{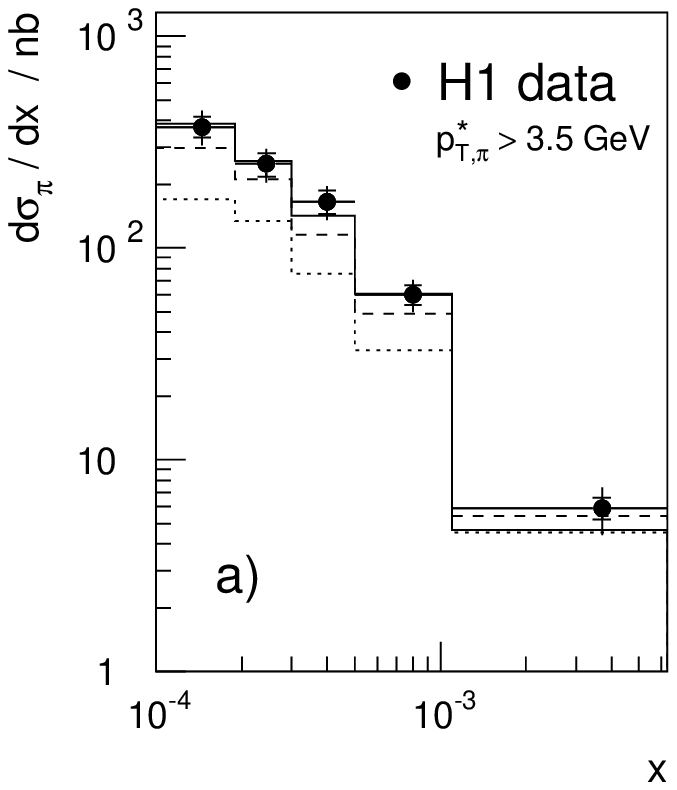}
\includegraphics{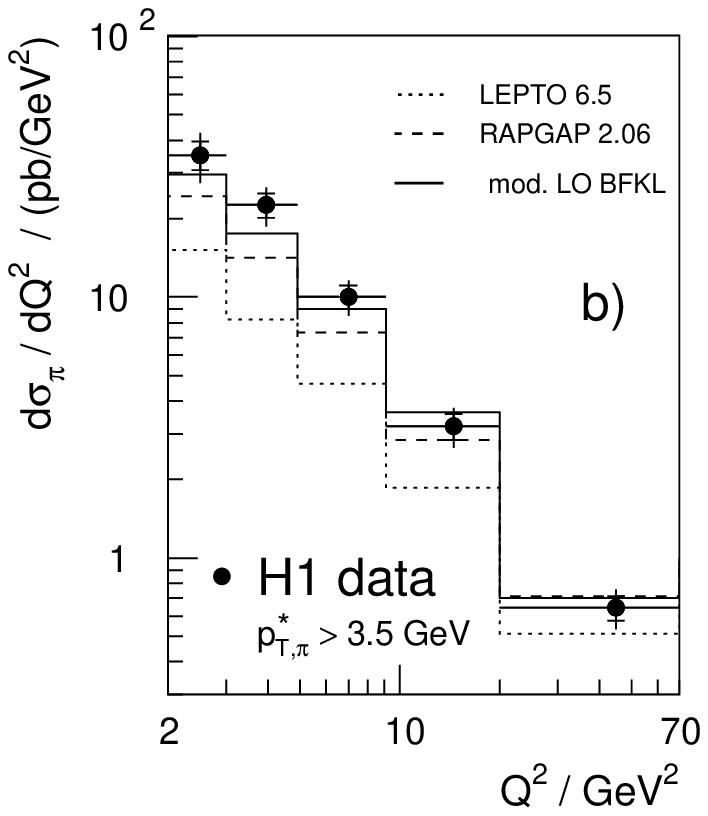}
\end{center}
\caption{Inclusive {\piz}-meson production cross-sections as a
  function of {\xbj} (a) and {\qsq} (b) for {\ptpi} $> 3.5$~GeV. 
  The phase space is given by 2.0 $<$ {\qsq} $<$ 70.0~{\gevsq},
  0.1 $<$ {\ybj} $<$ 0.6, $5^\circ <$ {\thpi} $< 25^\circ$ and 
  {\xpi} $=$ {\epi}$/${\eprot} $>$ 0.01.
   The inner error bars are 
statistical, and the outer error bars give the 
statistical and systematical error added quadratically.
  {\thpi} and {\xpi} are measured in the H1
  laboratory frame, {\ptpi} is calculated in the hadronic CMS. The
   QCD models RAPGAP (sum of direct and resolved contribution) and LEPTO
  are compared to the data. Also shown is the prediction of the LO
  BFKL calculation by Kwiecinski, Martin and Outhwaite.}
\label{fig:xbjq2b}
\end{figure}

\end{document}